\documentclass[prl,aps,twocolumn,showpacs]{revtex4}
\usepackage{graphics}
\begin{document}

\title{
Direct Optical Excitation of Quantum-Degenerate Exciton
States in Semiconductors}

\author{M.~Kira}
\author{S.W.~Koch}
\affiliation{
Department of Physics and Material Sciences Center,
Philipps-University Marburg, Renthof 5, D-35032 Marburg, Germany}
\author{G.~Khitrova}
\author{H.M.~Gibbs}
\affiliation{
Optical Sciences Center,
University of Arizona,
Tucson, AZ 85721, USA}
\date{\today}

\begin{abstract}
Quantum electrodynamic calculations predict that truly incoherent light
can be used to efficiently generate quantum-degenerate exciton population 
states. Resonant incoherent excitation directly converts photons into 
excitons with vanishing center of mass momentum. The populated exciton  
state possesses long-range order, is very stable against perturbations, 
and should be observable via its unusual 
directional and density dependence in luminescence measurements.   
\end{abstract}

\pacs{71.35.Lk, 03.75.Kk, 78.55.-m}

\newcommand{\be}{\begin{equation}}
\newcommand{\ee}{\end{equation}}
\newcommand{\bea}{\begin{eqnarray}}
\newcommand{\eea}{\end{eqnarray}}
\newcommand{\eps}{\varepsilon}
\newcommand{\ev}[1]{\langle#1\rangle}
\newcommand{\ddt}{\frac{\partial}{\partial t}}
\newcommand{\ihddt}{i\hbar\frac{\partial}{\partial t}}
\newcommand{\mcal}[1]{{\mathcal{#1}}}
\newcommand{\mrm}[1]{{\mathrm{#1}}}
\newcommand{\drm}{{\mathrm{d}}}
\newcommand{\e}{\mathrm{e}}
\renewcommand{\matrix}[1]{\mathbf{#1}}
\renewcommand{\vec}[1]{\mathbf{#1}}
\newcommand{\ci}[1]{\mathbf{#1}}
\newcommand{\stumm}{\bullet}
\newcommand{\ssum}[1]{\left[\left[#1\right]\right]}
\newcommand{\csum}[2]{\left|\left|#2\right|\right|_{#1}}
\newcommand{\widebar}[1]{\overline{#1}}

\maketitle
%
%\tableofcontents
%
%%%%%%%%%%%%%%%%%%%%%%%%%%%%%%%%%%%%%%%%%%%%%%%%%%%%%
%
%\baselineskip=1.5pc
%
%%%%%%%%%%%%%%%%%%%%%%%%%%%%%%%%%%%%%%%
%
Parallel to atomic Bose-Einstein 
condensation studies\cite{Bosecond}, semiconductor researchers 
are on an elusive and persistent quest\cite{Girardeau:59,Hanamura:77,Chase:79,Snoke:87,Lin:93,Butov:94,Butov:02,Snoke:02,Rapaport:04} 
to achieve macroscopic 
populations of quantum-degenerate
exciton states and Bosonic condensation. 
However, such attempts encounter several serious 
obstacles in GaAs-like direct band-gap systems since the elementary 
electron-hole pair excitations have a relatively short 
life time in the nanosecond range. Especially,
photoluminescence recombines very efficiently 
electron-hole pairs, and in particular 
low-momentum excitons, on a time scale 
of some tens of picoseconds, which leaves a strong hole-burning signature 
in the exciton center-of-mass momentum distributions.\cite{Kira:01} 
As a result, simple thermodynamic Bose-Einstein condensation 
is difficult to realize even for conditions where the large majority of 
electron-hole pairs exists in the form of excitons.
Thus, coupling to incoherent fields clearly seems to be a highly 
undesirable aspect for excitonic condensation 
since radiative exciton recombination opposes the macroscopic accumulation into 
the lowest energy and momentum state.
Consequently, most of the recent exciton-condensation 
experiments have either concentrated on 
${\rm Cu}_2{\rm O} $\cite{Snoke:87,Lin:93}
where the energetically
lowest exciton state is dipole forbidden, or on indirect semiconductor 
systems \cite{Butov:94,Butov:02,Snoke:02,Rapaport:04} 
with strongly suppressed radiative coupling.

In this context, it is an interesting question to
ask whether one can completely change the negative
influence of incoherent fields on condensation into a virtue 
{\it by using truly incoherent excitation pulses to pump --- not to drain --- 
direct band-gap systems.} To answer this question, we show in   
this Letter that by using quantum fields with finite intensity but 
vanishing phase, one can devise an incoherent optical excitation scheme
which selectively generates a very discrete set of low-momentum exciton states. 
This pumping directly seeds a many-body state with an
exciton population that exhibits long-range order. We show
that this state
is remarkably stable against Coulomb and phonon scattering and should
be directly observable experimentally via its directional emission and its
unusual dependence on excitation strength.  Furthermore, we discuss how
additional amounts of coherent excitation modify the
generation process. 

We investigate situations where a direct 
band-gap semiconductor is resonantly excited either with a fully or partially 
incoherent electromagnetic field. In our quantum-electrodynamic approach
we treat the light field via
bosonic photon operators $B_{{\bf q},q_\perp}$ and 
$B^\dagger_{{\bf q},q_\perp}$
related to a plane-wave mode with momentum $({\bf q},q_\perp)$,
where $\bf q$ and $q_\perp$ are the in-plane and 
perpendicular momentum components, respectively.
The semiconductor is assumed to be a two-band 
quantum-well or quantum-wire system,
described by fermionic operators $a_{c(v),{\bf}}$ and $a^\dagger_{c(v),{\bf}}$
related to destruction and generation of
conduction (valence) band electrons with in-plane momentum ${\bf k}$. 
We include the microscopic Coulomb interaction together with
coupling of carriers to photons and phonons.
The equations of motion for the interacting photon-carrier-phonon system
lead to the well-known hierarchy problem\cite{Haug:04}.
We use the cluster-expansion scheme\cite{cluster}
at a level where all two-particle correlations are fully included
while the three-particle correlations are treated at the scattering level.
As a result, the coherent part of the excitation is described by
the well-known Maxwell-semiconductor Bloch equations \cite{Haug:04} including
two-particle Coulomb and phonon scattering.\cite{Kira:04}

The most important novel aspect of the present studies 
is the modeling of the incoherent pumping
which, in contrast to coherent optical excitation, does not induce 
an interband polarization, 
but generates photon-assisted polarizations
$\Delta \langle
        B
        a^{\dagger}_{c} a_{v}
        \rangle$ 
($\Delta \langle
        B^\dagger
        a^{\dagger}_{v} a_{c}
        \rangle$)
describing correlated absorption (emission) of a photon while
an electron-hole pair is generated (recombined). These assisted
polarizations then act as sources for the generation of electron 
$f^e_{\bf k} \equiv \langle a^{\dagger}_{c,{\bf k}} a_{c,{\bf k}} \rangle$
and hole distributions
$f^h_{\bf k} \equiv \langle a_{v,{\bf k}} a^{\dagger}_{v,{\bf k}} \rangle$
as well as exciton correlations
$c_X^{{\bf q},{\bf k}',{\bf k}} \equiv 
\Delta\langle a^{\dagger}_{c,{\bf k}} a^{\dagger}_{v,{\bf k}'} 
a_{c,{\bf k}'+{\bf q}} a_{v,{\bf k}-{\bf q} }\rangle$
which are obtained from the two-particle expectation values
%$\langle a^{\dagger}_{1} a^{\dagger}_{2} a_{3} a_{4}\rangle$
by removing the single-particle contributions.
%$\langle a^{\dagger}_{1} a^{\dagger}_{2} a_{3} a_{4}\rangle_{\rm S}$.

From the semiconductor luminescence equations \cite{Kira:99}, we see that
the source terms for the photon-assisted polarizations are
%-----------------------
%-----------------------
\begin{eqnarray}
  &&\frac{\partial}{\partial t}
  \Delta \langle  B^{\dagger}_{{\bf q},q_\perp} 
  a_{v,{\bf k}-{\bf q}}^{\dagger} a_{c,{\bf k}} \rangle|_{\rm source}
  =
  {\cal F}_{q}
  [ 
          f^e_{{\bf k}} f^h_{{\bf k}-{\bf q}}
        +       
        \sum_{{\bf n}}
        c^{{\bf q},{\bf k},{\bf n}}_{\rm X}
  ]
\nonumber\\
  &&\;\;\;\;\;\;\;\;\;
  +(f^e_{{\bf k}} + f^h_{{\bf k}-{\bf q}}-1)
   \Delta \langle  B^{\dagger}_{{\bf q},q_\perp} 
   B_{{\bf q}_\Sigma} 
   \rangle. 
\label{eq:SLE_BY}
\end{eqnarray}
%-----------------------
%-----------------------
The self-consistent propagation of the incoherent light field is
computed from
%-----------------------
%-----------------------
\begin{eqnarray}
   &&\frac{\partial}{\partial t} 
   \Delta \langle B^{\dagger}_{{\bf q},q_\perp} 
   B_{{\bf q},q'_\perp} \rangle 
   = 
   i \left( \omega_{q} - \omega_{q'} \right) 
   \Delta \langle B^{\dagger}_{{\bf q},q_\perp}
   B_{{\bf q},q'_\perp} \rangle
\nonumber\\
   &&\;\;\;\;\;\;\;\;\;
   + {\cal{F}}_{q}
   \sum_{{\bf k}} 
   \Delta \langle B_{{\bf q}_{||},q'_\perp}
   a_{c,{\bf k}}^{\dagger} a_{v,{\bf k}-{\bf q}}\rangle 
\nonumber\\
   &&\;\;\;\;\;\;\;\;\;
   +{\cal{F}}^{\star}_{q'} 
   \sum_{{\bf k}}
   \Delta \langle B^{\dagger}_{{\bf q},q_\perp}
   a_{v,{\bf k}-{\bf q}}^{\dagger}  a_{c,{\bf k}} \rangle
\label{eq:SLE_b*b},
\end{eqnarray}
%-----------------------
%-----------------------
where ${\cal F}_q \equiv d_{vc} {\cal E}_{{\bf q},q_\perp} 
u_{{\bf q},q_\perp}/\hbar$, $d_{vc}$ is the dipole-matrix element of the 
direct band-gap semiconductor, ${\cal E}_{q}$ is the 
vacuum-field amplitude, and $u_{q}$ is the mode strength at the 
quantum well or wire position. Furthermore, we introduced the operator
$B_{{\bf q}_\Sigma} \equiv \sum_{q_\perp} d_{vc} 
{\cal E}_{{\bf q},q_\perp} u_{{\bf q},q_\perp} B_{\bf q,q_\perp}$ which
gives the total photon operator with momentum
component ${\bf q}$ along the confined system. 
Equation (\ref{eq:SLE_BY}) shows
that the photon-assisted polarization can be generated either 
via the stimulated term $\Delta \langle B^\dagger B \rangle$ or
via the spontaneous term including the electron-hole plasma 
source ($f^e f^h$) as well as the exciton-correlation contribution.

The incoherent source terms for the carrier distributions
and the excitonic correlations are 
%-----------------------
%-----------------------
\bea
  &&\hbar \frac{\partial}{\partial t} 
f^{e}_{\vec{k}}|_{\rm inc} 
= 
- 2{\rm Re}
  \left[
  \sum_{\vec{q}}
        \Delta \langle B^{\dagger}_{\vec{q}_\Sigma}\;
        a_{v,\vec{k} - \vec{q}}^{\dagger} a_{c,\vec{k}} 
        \rangle
   \right]
\label{eq:SBE-fe},
%%%%%%%%%%%%%%%%%%%%%%%%%%%%%%%%%%%%%%%%%%%%%
%%%%%%%%%%%%%%%%%%%%%%%%%%%%%%%%%%%%%%%%%%%%%
\\
  &&\frac{\partial}{\partial t} 
f^{h}_{\vec{k}} |_{\rm inc} 
=
- 2{\rm Re}
  \left[
  \sum_{\vec{q}}
        \Delta \langle 
        B^{\dagger}_{\vec{q}_\Sigma} 
        a_{v,\vec{k}}^{\dagger} a_{c,\vec{k} + \vec{q}}
        \rangle
  \right]
\label{eq:SBE-fh},
%%%%%%%%%%%%%%%%%%%%%%%%%%%%%%%%%%%%%%%%%%%%%
%%%%%%%%%%%%%%%%%%%%%%%%%%%%%%%%%%%%%%%%%%%%%
\\
  &&\hbar\frac{\partial}{\partial t}
  c^{\vec{q},\vec{k}',\vec{k}}_{\rm X}|_{\rm inc} 
=
 ( f^e_{\vec{k}}+f^h_{\vec{k}-\vec{q}}-1)
     \Delta \langle
        B^{\dagger}_{\vec{q}_\Sigma} 
        a^{\dagger}_{v,\vec{k}'} a_{c,\vec{k}'+\vec{q}}
        \rangle
\nonumber\\
 &&\;\;\;\;\;\;\;\;\;
  +(f^e_{\vec{k}'+\vec{q}}-f^h_{\vec{k}'}-1)
    \Delta \langle
        B_{\vec{q}_\Sigma} 
        a^{\dagger}_{c,\vec{k}} a_{v,\vec{k}-\vec{q}}
        \rangle
\label{eq:EXPcvcv}.
%%%%%%%%%%%%%%%%%%%%%%%%%%%%%%%%%%%%%%%%%%%%%
%%%%%%%%%%%%%%%%%%%%%%%%%%%%%%%%%%%%%%%%%%%%%
\end{eqnarray}
%-----------------------
%-----------------------
The full equation for 
$\frac{\partial}{\partial t}
  c^{\vec{q},\vec{k}',\vec{k}}_{\rm X}$ 
contains additionally the Coulomb and phonon terms responsible for 
the formation of exciton
populations including the true fermionic 
substructure.\cite{Usui:60,Kira:01,Hoyer:03} 

Since the ${\bf q}$ component of the photon is  
very small compared to any typical carrier momentum,
the exciton correlations couple to the incoherent
light field only when their center-of-mass momentum ${\bf q}$ 
is nearly vanishing. Experimentally, the in-plane photon momentum
${\bf q}$ can be fixed by controlling the excitation
direction which, e.g., for normal incidence leads to the exclusive population of 
the zero momentum exciton state. Similar selectivity clearly does not exist for
the carrier densities. 

In our numerical evaluations,
we assume pulsed incoherent excitation by choosing the initial
condition according to
%-----------------------
%-----------------------
\begin{eqnarray}
   \Delta \langle B^{\dagger}_{{\bf q},q_\perp} 
   B_{{\bf q},q'_\perp} \rangle|_{t=0} 
   = \Delta{}I_{\bf q}
	e^{-\left[(q_\perp -q_0)^2+(q'_\perp -q_0)^2\right]/\Delta q^2}
\label{eq:b*b_initial},
\end{eqnarray}
%-----------------------
%-----------------------
containing $\Delta{}I_{\bf q}$ which determines the incoherent
intensity in a given direction. The central frequency of the 
incoherent excitation
is $\omega_0 =c q_0$ while its energy and temporal width is determined
by $\Delta q$. Solving our coupled equations with only this incoherent
excitation source, we verify that no semiclassical optical polarization
is generated, however, the photon-assisted polarization 
builds up via the stimulated term in Eq.~(\ref{eq:SLE_BY}). 
This then leads to the generation of carrier distributions
in a wide range of
momentum states but exciton correlations 
are generated only in the low-momentum states
defined by $\Delta{}I_{\bf q}$. 

To illustrate the main effects of resonant excitation with truly
incoherent light, we numerically solve
the full set of equations (\ref{eq:SLE_BY})-(\ref{eq:EXPcvcv}) including
the semiconductor luminescence and Bloch equations  
with the microscopic Coulomb and phonon scattering included. 
As an example, we investigate the
case of a quantum-wire system with typical GaAs material parameters 
providing 11~meV exciton binding energy and a 3D-Bohr-radius of 
$ a_0 = 12.5$~nm.
We study low-temperature, resonant excitation configurations
such that only acoustic phonons 
need to be included. The combined treatment of the semiconductor
luminescence and Maxwell-semiconductor Bloch equations allows us to freely alter 
the coherence level of the excitation pulses since we can adjust
the relative intensities of the incoherent and coherent parts.
The generated many-body state after each excitation condition
can be followed by computing exciton distributions 
$\Delta n_\lambda({\bf q}) \equiv
	\sum_{{\bf k},{\bf k}'} 
	\phi^{\star}_\nu({\bf k}) \phi_\nu({\bf k})
	c^{{\bf q},{\bf k}'-{\bf q}_h,{\bf k}+{\bf q}_e}_X$ including
the exciton wave functions $ \phi_\nu({\bf k})$ \cite{Hoyer:03}.
Spatial correlations and long-range order effects
are determined from the correlation function between 
two electron-hole pairs,
%-----------------------
%-----------------------
\begin{eqnarray}
   g_{\rm ord}({\bf r}) \equiv \Delta \langle P^\dagger(0) P({\bf r}) \rangle
   = \frac{1}{{\cal V}^2} \sum_{{\bf q},{\bf k}',{\bf k}}
                          c_{\rm X}^{{\bf k},{\bf k}',{\bf q}} 
			  e^{i{\bf q} \cdot {\bf r}}
\label{eq:g_ord},
\end{eqnarray}
%-----------------------
%-----------------------
which involves creation $P^\dagger({\bf r})$ and annihilation
$P({\bf 0})$ of an electron-hole pair at different positions while
${\cal V}$ is the quantization volume.

%%%%%%%%%%%%%%
%  Figure 1. %
%%%%%%%%%%%%%%%%%%%%%%%%%%%%%%%%%%%%%%%%%%%%%%%%%%%%%%%%%%%%%%%%%
\begin{figure}[h]
\center{\scalebox{0.52}{\includegraphics{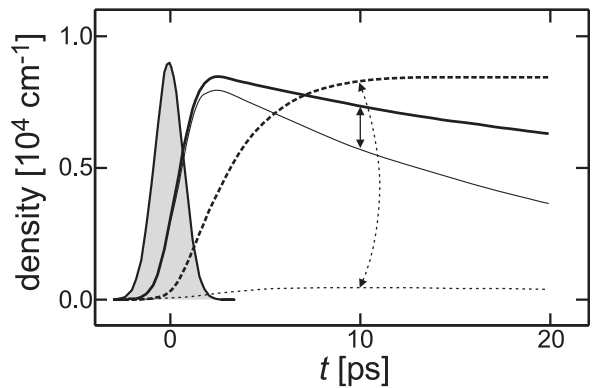}}}
\caption{
Generation of excitons with completely incoherent (solid lines) 
or coherent excitation (dashed lines) pulses for a temperature of $4K$.
The temporal evolution of the excitation pulse
(shaded area), the generated 1$s$-exciton density (thick lines),
and the density in the lowest momentum states (thin lines)
are shown. The arrows denote
curves belonging to the same excitation conditions.
} \label{fig1}
\end{figure}
%%%%%%%%%%%%%%%%%%%%%%%%%%%%%%%%%%%%%%%%%%%%%%%%%%%%%%%%%%%%%%%%%

For excitation at the 1$s$ exciton resonance, we show in
Fig.~\ref{fig1} a comparison of the generated incoherent exciton
density $\Delta{}N_{1s} = \frac{1}{\cal V} \sum_{\bf q} \Delta n_{1s}({\bf q})$
(thick solid and dashed lines) and the corresponding exciton density 
at the lowest momentum states (thin solid and dashed lines)
for both completely coherent 
(dashed line) and completely incoherent (solid line) excitation 
pulses of  the same intensity (dark shaded area). 
Even though both excitations give rise to well-above 90\% exciton fractions
with respect to the total carrier density 
$\frac{1}{\cal V} \sum_{\bf k} f^{e(h)}_{\bf k}$,
the coherent and incoherent excitations are characterized by very different
dynamics and final exciton states.
For the coherent case, the excitation first generates optical polarization 
which gradually is converted into incoherent 
exciton populations \cite{Kira:04}. However, 
less than $5\%$ of these 
excitons are found in the low-momentum states (thin dashed line).
In contrast, for completely incoherent excitation,
the direct conversion of photons into excitons leads to
an almost instantaneous build-up of a low-momentum exciton population 
(solid line). The generation of such a state is followed by 
radiative recombination of low-momentum excitons.

\begin{figure}[h]
\center{\scalebox{0.42}{\includegraphics{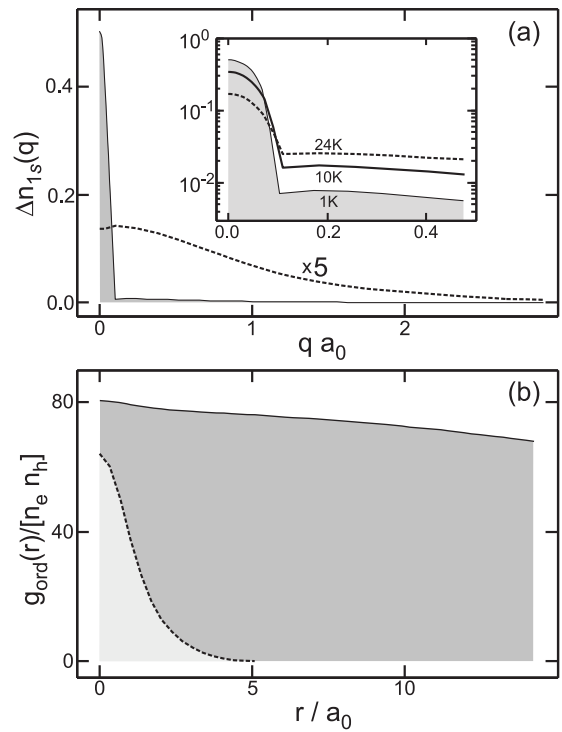}}}
\caption{
(a) Exciton distributions resulting from coherent 
(dashed line, curve multiplied by 5) 
and incoherent (shaded area) pumping for a lattice temperature $T=4K$
of Fig.~\ref{fig1} at time indicated by arrows.
The inset shows the incoherently generated distributions
at lattice temperatures of $1K$ (shaded area), $10K$ (solid line), 
and $24K$ (dashed line).
(b) The corresponding pair-correlation functions determined from 
Eq.~(\ref{eq:g_ord}).
} \label{fig2}
\end{figure}

In order to gain more insight into the incoherent excitation
process, we compare the generated exciton distributions for  
coherent and incoherent excitation conditions. For this purpose, we determine
$\Delta n_{1s}({\bf q})$ at a time moment
when the generation process has been completed
(indicated by arrows in Fig.~\ref{fig1}). Figure \ref{fig2}a
shows the incoherently (shaded area) and 
coherently (dashed line) generated $\Delta n_{1s}({\bf q})$
determined at 10~ps after the pulse maximum for a lattice temperature $T=4K$. 
We observe that the coherently 
generated populations have a broad momentum distribution, whereas
the incoherent excitation leads to a highly singular 
1$s$-exciton distribution around the zero-momentum state. This 
distribution is remarkably stable against carrier and phonon
scattering (see inset to Fig. \ref{fig2}a). 
In Fig.~\ref{fig2}b we plot the 
corresponding pair-correlation functions showing that
for the coherently generated state the pair-correlation function 
$g_{\rm ord}({\bf r})$ (dashed line) decays on the length scale 
of the 3D-Bohr radius $a_0$, whereas the incoherent 
$g_{\rm ord}({\bf r})$ (shaded area) exhibits pronounced long-range 
order. The observation of 
{\it the singular $1s$-exciton distributions and the existence of 
long-range order uniquely demonstrate that incoherent pumping 
directly produces a highly quantum-degenerate exciton state.}

\begin{figure}[h]
\center{\scalebox{0.54}{\includegraphics{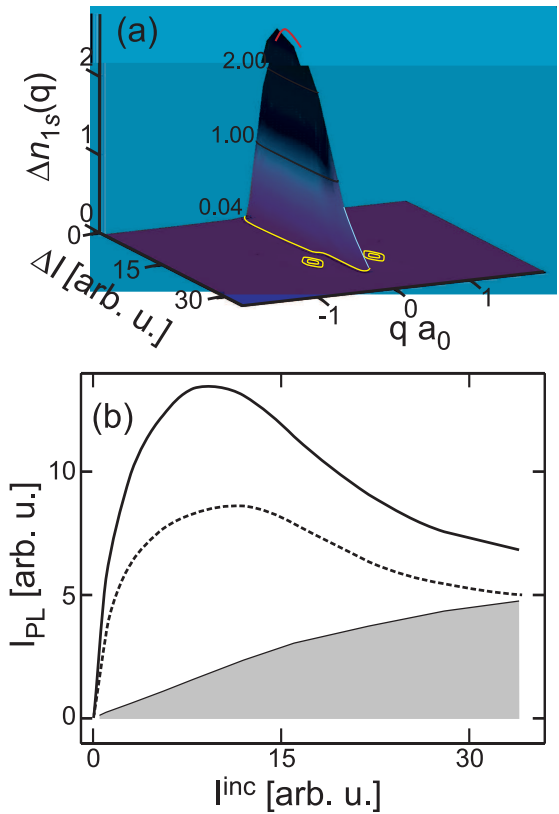}}}
\caption{
(a) Exciton distributions as function of intensity of 
incoherent field determined 16~ps after the excitation
for the 100\% incoherent excitation at $T=4K$.
(b) Total photoluminescence
resulting from 100\%ccoherent (shaded area), 40\%coherent (dashed line), 
and 100\%incoherent (solid line) pumping determined 16~ps after 
the excitation.
} \label{fig3}
\end{figure}

Now we turn to the questions concerning the robustness of the incoherently
generated exciton state
and how to observe it in experiments. In Fig.~\ref{fig3}a,
we plot the exciton population for different intensities of the 
incoherent field showing that the macroscopic population of the $1s$ state 
continuously increases up to the intensity level $\Delta{}I = 10$,
which corresponds 
to a generated carrier density of $n a_0 = 0.1$. For this situation, 
basically all of the excitons are in the low-momentum state. 
For higher excitation levels, the $1s$-exciton population starts
to decrease because the underlying fermion character of the 
electron-hole pairs gradually prevents further exciton accumulation. 
The quantum-degenerate state 
ceases to exist above  
$\Delta{}I = 31$ corresponding to a
density $n_e a_0 = 0.3$. In our studies of lattice temperature effects,
we find that the results presented here are
valid for temperatures of less than 
approximately $6 - 10K$. For higher temperatures, acoustic phonon
scattering gradually leads to population scattering into
higher momentum states (see inset to Fig.~\ref{fig2}a). 

%%%%%%%%%%%%%%
%  Figure 2. %
%%%%%%%%%%%%%%%%%%%%%%%%%%%%%%%%%%%%%%%%%%%%%%%%%%%%%%%%%%%%%%%%%
%%%%%%%%%%%%%%%%%%%%%%%%%%%%%%%%%%%%%%%%%%%%%%%%%%%%%%%%%%%%%%%%%

Since we investigate the case of direct-gap 
semiconductors, the macroscopic $1s$-exciton population 
recombines at the rate of the radiative decay time. The 
corresponding photoluminescence is emitted with the same 
in-plane momentum as the condensate. As a result, the presence
of a macroscopic population at ${\bf q} = {\bf 0}$ should show up 
as a strong directional dependence of the luminescence.\cite{Keeling:04}

%%%%%%%%%%%%%%
%  Figure 3. %
%%%%%%%%%%%%%%%%%%%%%%%%%%%%%%%%%%%%%%%%%%%%%%%%%%%%%%%%%%%%%%%%%
%%%%%%%%%%%%%%%%%%%%%%%%%%%%%%%%%%%%%%%%%%%%%%%%%%%%%%%%%%%%%%%%%

Additionally, the intensity dependence of the population in
Fig.~\ref{fig3}a manifests itself as an unusual intensity dependence of  
the integrated photoluminescence,
$I_{\rm PL} = \sum_{\bf q}
\frac{\partial}{\partial t} 
\Delta \langle B^\dagger_{\bf q} B_{\bf q} \rangle$. 
As an example,
we show in Fig.~\ref{fig3}b the variation of 
$I_{\rm PL}$ as function of pump intensity
determined 16~ps after 100\% incoherent
(solid line), 40\% coherent (dashed line), and 100\% coherent
(shaded area) excitation.
We see that for the relatively low levels of excitation
studied here, the total luminescence for the 100\% coherent case exhibits 
the expected practically linear dependence on the excitation strength. However, when
we include a significant incoherent component to the excitation process,
 $I_{\rm PL}$ behaves strongly nonmonotonously. For 100\% incoherent
excitation, the luminescence is maximized at the intensity level 
$\Delta{}I = 10$ corresponding to the maximum singularity of
exciton distributions in Fig.~\ref{fig3}a. Since the zero momentum state
population decreases for elevated intensities, the
luminescence decreases also until it reaches the same level
as that for coherent excitation. This predicted distinct
nonmonotonic behavior of $I_{\rm PL}$ should be directly observable
in experiments serving as a clear signature
for the formation of the quantum-degenerate exciton state. 
We note in Fig.~\ref{fig3}b, that $I_{\rm PL}$ has a 
maximum even in the presence of 40\% coherent excitation, indicating that a
significant population in the quantum-degenerate state 
is generated even in this imperfect case.

In summary, our microscopic calculations lead us to predict
that resonant incoherent excitation with fields of
finite intensity but vanishing phase directly generates
a macroscopic population of 1s-excitons 
in low-momentum states. This quantum-degenerate
state possesses significant long range order and is 
remarkably stable against perturbations. 
As predicted experimental signatures, the integrated luminescence 
displays distinct non monotonic behavior since it 
first increases and then decreases with increasing excitation intensity.

\begin{acknowledgments}
The Marburg work is funded by the Optodynamics Center and
the Deutsche Forschungsgemeinschaft through
the Quantum optics in semiconductors research group.
The Tucson work is funded by NSF AMOP.
\end{acknowledgments}

\bibliography{myrefs}

\end{document}